\documentclass[letterpaper,twocolumn,fleqn]{article} 

\usepackage{ist}
\usepackage{multirow}
\usepackage{mwe}
\usepackage{subcaption}
\usepackage{subfiles}
\usepackage{times}
\usepackage{epsfig}
\usepackage{graphicx}
\usepackage{amsmath}
\usepackage{amssymb}
\usepackage{cleveref}

\usepackage[style=ieee]{biblatex}
\title{A Visual Quality Assessment Method for Raster Images in Scanned Document}
\author{Justin Yang$^{*}$, Peter Bauer$^{\dagger}$, Todd Harris$^{\dagger}$, Changhyung Lee$^{\ddagger}$, Hyeon Seok Seo$^{\ddagger}$, Jan P Allebach$^{*}$, Fengqing Zhu$^{*}$;\\
$^{*}$Elmore School of Electrical and Computer Engineering, Purdue University, West Lafayette, Indiana, USA\\ $^{\dagger}$HP Inc., Boise, Idaho, USA\\ $^{\ddagger}$HP Printing Korea Co Ltd, Suwon, Korea}

\date{} % date has an empty field.

\begin{document} 

\maketitle 

\thispagestyle{empty} % prevents the first page to be numbered

%%%%%%%%%%%%%%%%%%%%%%%%%%%%%%%%%%
% Abstract
%%%%%%%%%%%%%%%%%%%%%%%%%%%%%%%%%%

\begin{abstract}
\label{sec:abstract}
Image quality assessment (IQA) is an active research area in the field of image processing. Most prior works focus on visual quality of natural images captured by cameras. In this paper, we explore visual quality of scanned documents, focusing on raster image areas. Different from many existing works which aim to estimate a visual quality score, we propose a machine learning based classification method to determine whether the visual quality of a scanned raster image at a given resolution setting is acceptable. We conduct a psychophysical study to determine the acceptability at different image resolutions based on human subject ratings and use them as the ground truth to train our machine learning model. However, this dataset is unbalanced as most images were rated as visually acceptable. To address the data imbalance problem, we introduce several noise models to simulate the degradation of image quality during the scanning process. Our results show that by including augmented data in training, we can significantly improve the performance of the classifier to determine whether the visual quality of raster images in a scanned document is acceptable or not for a given resolution setting.
\end{abstract}

%%%%%%%%%%%%%%%%%%%%%%%%%%%%%%%%%%%%
% Overall Document Guidelines: Head
%%%%%%%%%%%%%%%%%%%%%%%%%%%%%%%%%%%%
\section{Introduction}
\label{sec:intro}
Scanners on mulit-functional printers (MFPs) are widely used for both personal and commercial purposes to digitize printed documents. The scan resolution is usually set by the user or by the default setting in the scanner. This leads to the same output file size regardless of the content of the scanned document. 
% the fact that when scanning, disregard of the image content of the whole page, the final output file size will be the same. 
However, most printed documents do not require scanning all regions in one page at a high resolution setting. This can lead to a unnecessarily large file size for those regions where details are less relevant, such as images that do not contain salient objects or are mostly homogeneous. On the other hand, when scanning at a lower resolution, for image regions where details are important, information could be lost and visual quality degraded. Striking a good balance between visual quality and compression level to optimize user experience is a valuable yet challenging problem since it depends not only on the content of the document but also the purpose of scanning the document. For a scanned document that contains multiple raster image regions on the same document, the best strategy is to scan the raster image regions with different resolutions according to the individual image content in order to maintain desired image quality with minimum file size. 

We are interested in developing methods that would be implemented into MFPs, which uses only the CPU to perform computation. Given this setup, we need to design an efficient image quality assessment method that can balance the performance and complexity suitable for product deployment. Previous works on document image quality assessment focus on Optical Character Recognition (OCR) accuracy \cite{DIQ_DL_Le, DIQ_Kanungo} instead of visual pleasantness. Other works on optimal resolution determination for scanned documents mainly focus on the text regions of a scanned document \cite{litao_IQA,litao_TextQA} such as bar-codes/QR-codes, hand-writings, and printed text. In these methods, the focus is on determining suitable resolution to achieve certain level of accuracy in OCR, and treat OCR accuracy as the metric to determine the visual quality of the document. However, in our scenario, we want to focus on raster image content, which does not have such metric to utilize for quality assessment. Since there is no objective evaluation more suitable than the human eyes, a psychophysical experiment is needed to determine the visual acceptance of raster images in scanned documents.

Due to the limited computation power, we propose to extract a set of features from the scanned document that are relevant to visual quality, require relatively simple computation for CPUs, and then find the appropriate combinations. Under a simple machine learning based framework, we then fuse the information from different features to determine the minimum required scan resolution. 
% The selected metrics/features should require relatively simple computation for CPUs. 
% We propose to combine multiple features to determine the minimum required resolution because each feature contains visual quality information from different aspects. By using machine learning techniques, we are able to fuse the information from different features and then properly combine them to achieve the desired performance.  

The main contributions of this paper are as follows:
\begin{itemize}
    \item We formulate visual quality assessment for document images into full-reference task, as it is different from natural camera image quality assessment where no reference image can be provided.
    \item We propose a simple machine learning method to estimate image quality for different image regions in a scanned document page to determine the optimal scanning resolution for each image region. 
    \item We collect a scanned document dataset containing images scanned at different resolution settings paired with human subjective ratings from a psychophysical experiment.
    \item We address the data imbalance problem of subjective ratings by introducing noise models that simulate the degradation of image quality during the scanning process.  
\end{itemize}
% 1. We have collected a dataset containing pairs of human subjective ratings and image with different resolution settings
% 2. We dealt with the data imbalance issue of subjective ratings by introducing noise model that simulate the degradation of image quality during the scanning process.  
% 3. We proposed a model based on SVM to estimate image quality for different image regions in a single scanned document which can further infer the optimal scanning resolution for each image region. 

% The remainder of the paper is organized as follows. In Section~\ref{sec:related}, we review prior works related to image quality assessment, and optimal resolution determination for scanned documents. In Section~\ref{sec:datacollection}, we introduce the collection of a scanned documents dataset including the dataset preparation and composition. In Section~\ref{sec:IQA}, a detailed description of the psychophysical experiment is provided. Our proposed method to determine the optimal scanned document resolution is described in Section~\ref{sec:method}. Section~\ref{sec:dataanaysis} summarizes the statistics of the psychophysical experiment and introduces three noise models to address the data imbalance problem from the human subject ratings. Evaluation of proposed method and discussion are presented in Section~\ref{sec:result}. Section~\ref{sec:conclusion} presents a conclusion of our work.

% \section{Related Works}
% \label{sec:related}
% \subfile{sections/02_Related_Works.tex}

\section{Dataset Collection}
\label{sec:datacollection}

\subsection{Composition of Our Dataset}
\label{Composition_Of_Dataset}
\begin{figure}[t]
  \centering
  \includegraphics[scale=0.4]{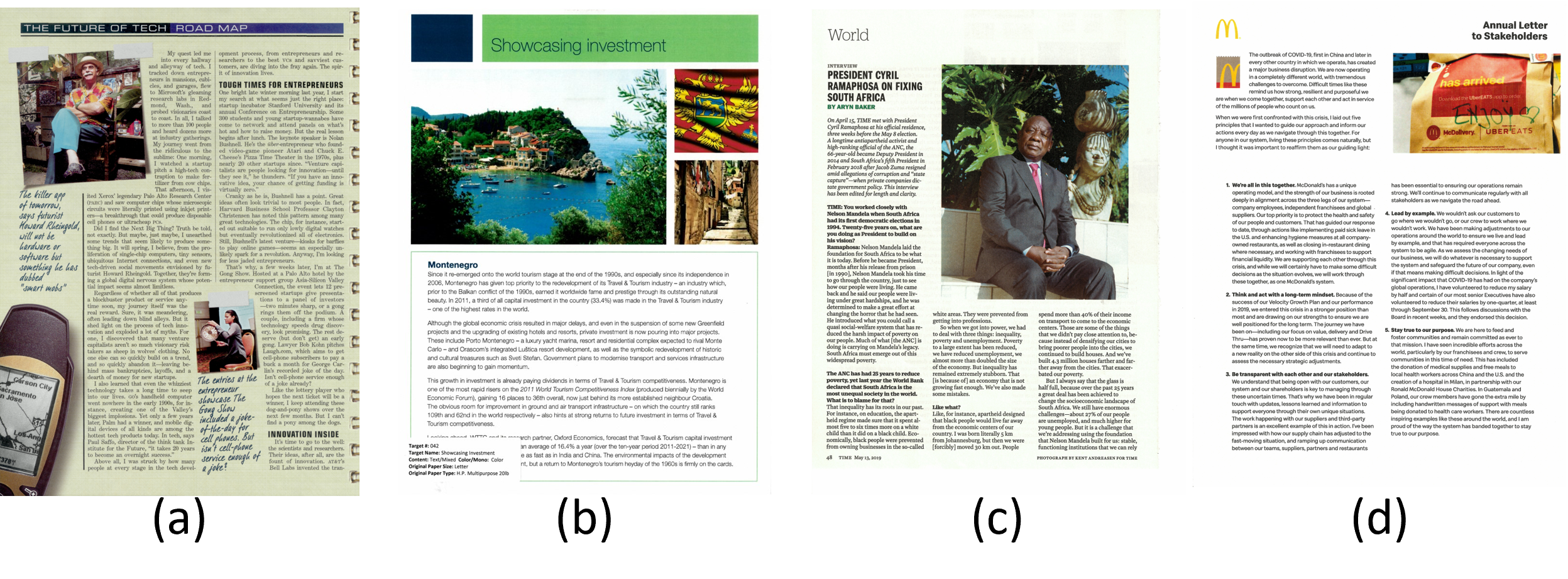}
    \caption{Samples from each subset of the collected dataset. (a) - (d) correspond to an example from the Prima dataset, MRC dataset, MS dataset, and HQS dataset, respectively.}
  \label{fig:data}
\end{figure}
We collected a dataset of mixed content scanned documents containing raster images and possibly other elements such as text, barcodes, line art, graphs etc.
We grouped the documents into 4 subsets: Prima Dataset \cite{Prima_Paper},  Mixed Raster Content (MRC) Dataset, Magazine Scans (MS), and High Quality Scans (HQS). The Prima dataset \cite{Prima_Paper} is a public dataset which contains documents, magazines, and technical journals that are scanned at 300 dpi. The MRC subset is an internal dataset from HP Inc. This includes documents with mixed raster content scanned at 300 dpi. For the MS subset, it consists of 300 dpi scans from Time Magazine using the HP OfficeJet MFP X585.  The HQS subset contains high quality document sources from the internet, which includes annual reports from companies and universities. These documents were first printed using an HP LaserJet 500 color MFP M575 and then scanned at 300 dpi using an HP OfficeJet MFP X585. 
Figure \ref{fig:data} shows some examples from each component of our dataset. For our dataset, we use 50 documents from the Prima dataset, which contains a total of 105 raster images; 5 from MRC dataset, which contains 7 raster images; 5 documents from the MS dataset, which contains 5 raster images; and 40 documents from the HQS dataset, which contains 57 raster images. For the Prima dataset, there are 1 to 4 raster images in each page. For the other subsets of our dataset, there are 1 to 2 raster images in each page.

\subsection{Generation of different dpi images}
\begin{figure}
  \centering
   \includegraphics[width=1\linewidth]{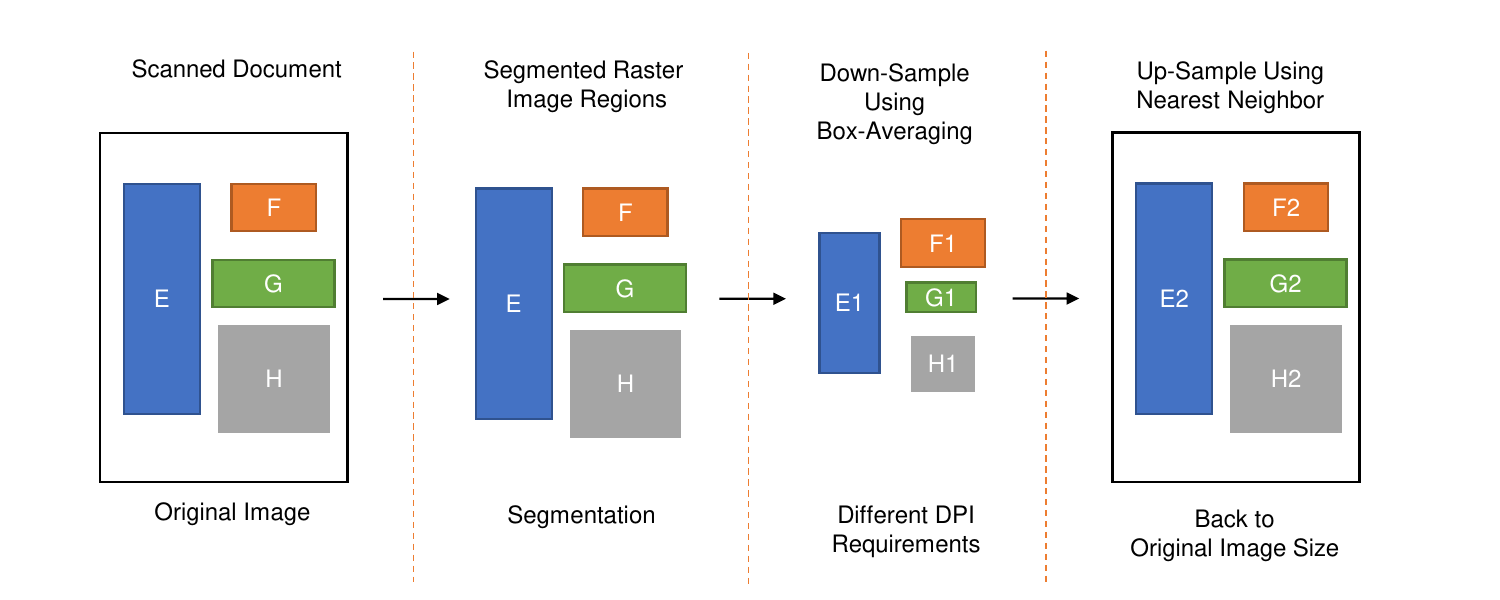}
   \caption{Pipeline for generating different dpi images.}
   \label{fig:downsample_diagram}
\end{figure}
% In this subsection, we will first describe the overall pipeline of our system. 
Given a mixed content document page scanned at 300 dpi, which we term as the base resolution, our goal is to determine the optimal resolution setting for each individual raster image region in this document. Figure \ref{fig:downsample_diagram} shows an example of this process. Assume the original image consists of 4 regions, namely E, F, G, and H.  Each region may a have different optimal scan resolution setting. In our system, segmentation is first applied to the full scanned document to separate each region (\textit{e.g.}, text region or image region).  We manually label the segmentation map using the LabelMe tool \cite{wada2018labelme}.  

To determine the optimal scan resolution of a given raster image region in a document page, we generate low dpi versions of each region in addition to having the base resolution, which is then assessed by our proposed method (described in Section \ref{sec:method}).
% Then the optimal resolution is assigned based on the raster image content .  
In order to view the down-sampled image on the original scanned document, region-based post-processing is applied to the images to emulate the visual effects of low-dpi images. Two steps are included in this process to generate the low dpi images, the down-sampling phase and the up-sampling phase. Depending on the chosen optimal resolution, if it is other than base resolution, \textit{i.e.,} 300 dpi, box averaging is applied to the original image region. This is called the down-sampling phase. After the down-sampling phase, the decimated image is then interpolated back to the same size as the original image, termed the up-sampling phase which uses the nearest neighbor algorithm.

\vspace{-0.1cm}
\section{Psychophysical Image Quality Assessment Experiment}
\label{sec:IQA}

To determine the minimum resolution for our purpose, we want to minimize the dots per inch (dpi) resolution setting for a multifunction printer (MFP) when scanning a document page. Our goal is to reduce the resolution setting as much as possible without noticeable visual quality degradation, so as to meet the end user's expectations. We design a psychophysical image quality evaluation experiment as the subjective quality metric to compare the visual quality between different dpi settings of the raster images in scanned documents.

The experimental procedure is as follows. In order to keep the experiment in a reasonable time length, we first partition our dataset into five subsets so that each participant only need to rate one subset instead of all images in the dataset. The size of the whole document would be set at 8.5 by 11 inches for display. There were a total of 35 participants with experience in the image processing field.
% which included four different HP engineers with background in image quality assessment, two professors with background in image processing and computer vision, and 29 graduate students 
For each given mixed raster document, the participant can select the raster images one-by-one, and set the resolution of that image areas to one of the following: 100, 150, 200, and 300 dpi, while the remaining raster images on the page, \textit{i.e.}, the text and vector regions, are fixed at 300 dpi, which is the base resolution setting. Participants are asked to rate each raster image individually in the document according to the naturalness, smoothness and detailed rendition of the raster image area of the document. Participants can select from four quality scores following the standard mean opinion score (MOS) setting for quality assessment: A (Visually Pleasant), B (Visually Okay), C (Visually Okay with Some Artifacts), and D (Visually Unacceptable). Each participant rates approximately 30 separate raster images in 15 to 20 pages, where each page contains 1 to 4 raster images. The whole procedure took on average 25 minutes to complete. Each raster image in the scanned document is evaluated by seven different participants, and assigned a quality rating. The final quality rating assigned to each raster images is calculated by averaging the ratings ($A=4$, $B=3$, $C=2$, $D=1$) and then quantize the averaged result to the closest category after removing outliers.

%Figure \ref{fig:GUI_example} shows an example of the GUI application used by the participants to provide quality scores. The window on the left shows the entire document page. The right side of the GUI contains a separate column for each raster image on the page. Clicking the box in the second row highlights the corresponding raster image in the full page display. Then, by clicking the boxes next to each resolution in that column, the selected raster image is re-rendered at that resolution following the procedure described earlier; and the participant can select A, B, C, or D from the drop-down menu to rate that raster image at the given resolution.
\vspace{-0.1cm}
\section{Method}
\label{sec:method}
\subsection{Overview of the Method}
\begin{figure}[h]
  \centering
   \includegraphics[width=1\linewidth]{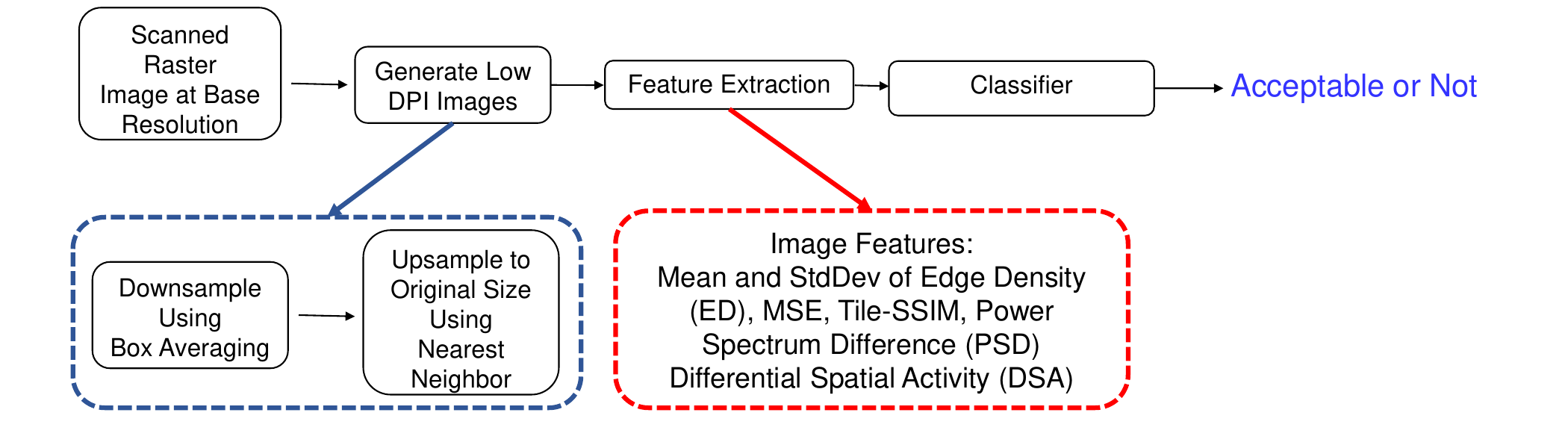}
   \caption{Overview of the method.}
   \label{fig:overview_diagram}
\end{figure}

In this section, we describe the proposed method to estimate image quality for different image regions in a scanned document page which can be used to determine the optimal scanning resolution for each image region.
Our system consists of two phases: the training phase and the inference phase. During the training phase, we utilize the ground truth ratings collected from the psychophysical experiment and several full-reference image features to train a classifier that predicts raster image quality. We formulate the research question into a binary classification problem by categorizing quality scores A (Visually Pleasant) and B (Visually Okay) into one class as ``Visually Acceptable", and categorizing C (Visually Okay with Some Artifacts) and D (Visually Unacceptable) into a second class as ``Visually Unacceptable". The multi-choice setting in ratings enables a more flexible threshold for different applications with different requirements. For example, in business applications, if higher quality is required, then we can categorize only quality scores A as "Visually Acceptable" and the rest as "Visually Unacceptable". In this paper me mainly focus on the standard setting, which is A,B as "Visually Acceptable" and C,D as "Visually Unacceptable". A Support Vector Machine \cite{SVM} is used to fuse different input image features and take joint consideration for training. At the inference stage, given an image pair including the original document scanned at the base resolution dpi, \textit{e.g.}, 300 dpi, and the lower resolution image, we first extract the set of image features. The feature vectors are then merged and are fed to the trained support vector machine model to classify if the given raster image quality is acceptable or not to derive the minimum acceptable resolution setting. An overview of the proposed method is illustrated in Figure \ref{fig:overview_diagram}.

\subsection{Feature Extraction}
For no-reference metrics, we include mean and stddev Edge Density (ED), which is the Canny edge map \cite{canny1986computational} intensity for the low resolution images. For full-reference metrics, the low resolution images are resized back to original scale of base resolution. The features we used in our method include Differential Spatial Activity (DSA), the root mean square difference between Sobel edge maps of the two images \cite{DSA}; mean and stddev Mean Square Error (MSE); mean and stddev Power Spectrum Difference (PSD), difference of 2D discrete-time Fourier transform images between the two images, and the mean and stddev of Tile-SSIM \cite{tile_ssim}. In order to calculate Tile-SSIM, we follow the method in \cite{tile_ssim}. 
For the mean and standard deviation (stddev) for each feature, the different feature maps are partitioned into tiles, and then calculate the mean and stddev among different tiles. The corresponding tile size for 300dpi, 200dpi, 150dpi, 100dpi are 12, 8, 6, 4, respectively.%\ref{tab:scan_tile_table}. 

\subsection{Feature Selection}
Feature selection is the process of selecting a subset of relevant features from the original feature set. A good feature selection not only can compress the dimensionality of the feature space, but also improve the learning efficiency at the training stage and the prediction accuracy at the inference stage by preventing the trained model from overfitting. 
% After feature selection, the model size is typically smaller and more efficient. 
% Feature selection is commonly used in pattern recognition \cite{fs_for_pr}, data mining \cite{fs_for_dm}, etc., and can be designed based on statistics and information theory \cite{feature_selection_survey}. 

% Feature selection not only helps to improve the classification performance of the model, but also prevents it from overfitting. It also makes the model size smaller and more efficient. Finally it provides insight for the dataset that implies the importance of different feature. 
Our feature selection method follows the standard process of sequential floating forward selection (SFFS) \cite{SFFS}, which reduces an initial $t$-dimensional feature space, where $t$ is the total number of features, to a target $d$-dimensional feature subspace, $d<t$. After the feature selection process, the selected features are normalized, equally weighted and concatenated. The fused feature vector from the training image set is used to train a support vector machine (SVM) based classifier. The classifier makes prediction of whether the image quality is acceptable or not based on these features. 

\section{Data Analysis and Augmentation}
\label{sec:dataanaysis}

\subsection{Noise Models}
Table \ref{tab:data_summary} summarizes the participant ratings from the psychophysical experiment. Because the quality scores resulted from our dataset is significantly unbalanced, \textit{i.e.}, more images are rated at ``Acceptable" compared to ``Unacceptable", we introduce several noise models to simulate the quality degradation during scanning process in order to mitigate the data imbalance issue. 

\begin{table}[t]
\caption{Subjective Rating Summary}
    \centering
    \begin{tabular}{|c|c|c|} 
        \hline 
        Resolution (dpi)& \shortstack{Data Rated as \\ Acceptable}& \shortstack{Data Rated as \\ Unacceptable}\\ \hline
        300&174&0\\ \hline
        200&173&1\\ \hline
        150&154&20\\ \hline
        100&76&98\\ \hline
    \end{tabular}
\label{tab:data_summary}
\end{table}

We adopt several noise models to augment visually unacceptable images to make the dataset more balanced, including Gaussian noise, salt and pepper noise, and generative adversarial networks. The parameter settings for the former two noise models are determined based on the Structural Similarity Index (SSIM) statistics of the images rated as ``Unacceptable", while the latter is trained using a subset of the images rated as ``Unacceptable". 

\begin{figure}[!h]
  \centering
  \includegraphics[scale=1]{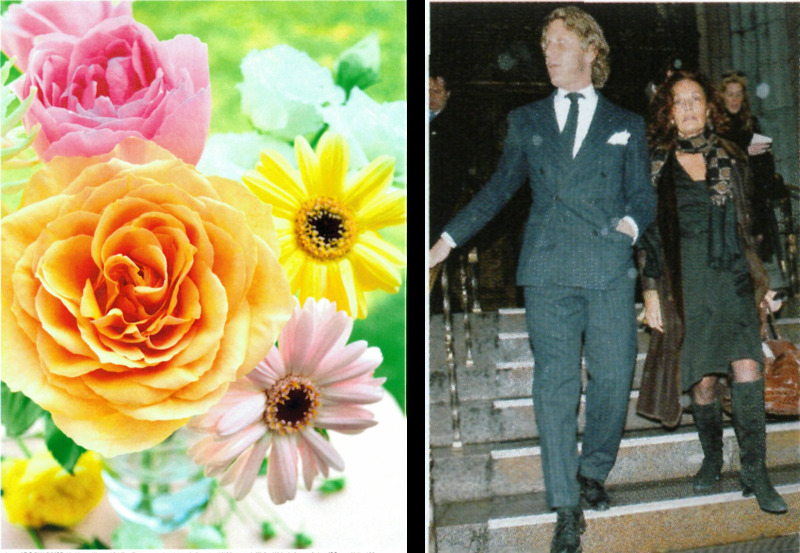}
    \caption{Example of high (left) and low (right) quality scans.}
  \label{fig:data_high_low_quality}
    \label{fig:data_high_low_quality}
\end{figure}

\textbf{Gaussian Noise.}
As we discovered that Gaussian noise can simulate the visual quality degradation effect of the scanned documents during the scanning process, we decided to use it as one of our augmentation methods. The variance of the Gaussian noise applied to images is set to $0.0005$ as it matches the mean SSIM value of the ``Unacceptable" images in our dataset. %Example images with and without Gaussian noise can be found in Figure \ref{fig:figure_gaussian}.

% \begin{figure}[!htbp]
% \centering
%     \subfigure
%         \centering
%         \includegraphics[width=0.45\linewidth]{images/flower_compare.png}
%     \subfigure
%         \centering
%         \includegraphics[width=0.45\linewidth]{images/flower_compare_gau.png}
%     \caption{Gaussian noise degraded dataset example: original example image (left) and the corresponding version degraded by adding Gaussian noise (right).}
% \label{fig:figure_gaussian}
% \end{figure}

\textbf{Salt and Pepper Noise.}
Impulse noise may be introduced during the document scanning process, which can be used simulate visual quality degradation. Therefore, We use salt and pepper noise to augment our dataset. The noise density of the salt and pepper noise applied to the images is set to $0.008$ in order to match the mean SSIM value of the ``Unacceptable" images in our dataset.% Example images with and without salt and pepper noise can be found in Figure \ref{fig:figure_salt}.
% \begin{figure}[!htbp]
% \centering
%     \subfigure
%         \centering
%         \includegraphics[width=0.45\linewidth]{images/girl_compare.png}
%     \subfigure
%         \centering
%         \includegraphics[width=0.45\linewidth]{images/girl_compare_s&p.png}
%     \caption{Salt and pepper noise degraded dataset example: an original example images (left) and the corresponding version degraded by adding salt and pepper noise (right).}
%     \label{fig:figure_salt}
% \end{figure}

\textbf{Generative Adversarial Network (GAN).}
Generative adversarial Networks, or GANs, are popular methods for image generation and synthesis \cite{NIPS2014_5ca3e9b1}. 
% A GAN consists of two main parts: the generator $G$ and the discriminator $D$. The main idea to train a GAN is to let the generator and discriminator play a mini-max game where the generator tries to synthesize a realistic looking image, while the discriminator tries to distinguish the real image from the synthesized one. %The loss function is described as below where $p_{data}(x)$ serves as the real data distribution while the $p_z(z)$ represents a random noise distribution.
% \begin{equation}
%     \begin{array}{r}
%         \min _{G} \max _{D} V(G, D)=E_{x \sim P_{d a t a}(x)}[\log D(x)] \\
%         +E_{z \sim P_{z}(z)}[\log (1-D(G(z)))]
%     \end{array}
% \end{equation}
% where $G$ is the generator while $D$ is the discriminator. The generator maps a random noise $z$ into a the real data $X$ and results in the generated data $G(z)$. $D$ is a classifier that takes both the real data $X$ and the generated $G(z)$ to determine if $G(z)$ is real or not. When the sample comes from $G(z)$, $D$ would minimize the loss. When the sample comes from $X$, $D$ would maximize the loss.
For our noise model, we formulate the task of generating low quality images into an image-to-image translation task, where we utilize conditional GANs \cite{NIPS2014_5ca3e9b1} to generate low quality versions of the given images. The loss function is slightly different from the original GAN's loss function as we also take the base resolution image $X$ as input for our network. The loss function is described as below.
\begin{equation}
    \begin{array}{r}
        \min _{G} \max _{D} V(G, D)=E_{x \sim P_{data}(x)}[\log D(y)] \\
        +E_{z \sim P_{z}(z)}[\log (1-D(G(x,z)))]
    \end{array}
\end{equation}
where $G$ is the generator and $D$ is the discriminator. The generator takes the base-res image $x$ and a random noise $z$ as input and generates the low-res data $G(x, z)$. $D$ is a classifier that takes both the real low-res data $y$ and the generated $G(x, z)$ as inputs to determine if $G(x, z)$ is real or not. When the sample comes from $G(x, z)$, $D$ would minimize the loss. When the sample comes from $y$, $D$ would maximize the loss.
% \begin{figure}[h]
% \centering
%     \subfigure
%         \centering
%         \includegraphics[width=0.45\linewidth]{images/man_compare.png}
%     \subfigure
%         \centering
%         \includegraphics[width=0.45\linewidth]{images/man_compare_GAN.png}
%     \caption{Conditional GAN degraded dataset example: an original example images (left) and the corresponding version degraded using conditional-GAN.}
% \label{fig:figure_gan}
% \end{figure}
%An example original image and a degraded version of that image generated with a conditional GAN can be found in Figure \ref{fig:figure_gan}.

\subsection{Analysis on Augmented data}
To determine the noise level used in the data augmentation for each noise model, we derive statistics from the images that are labeled as ``Unacceptable" from the psychophysical experiment. In particular, we calculate the mean and standard deviation of the SSIM (Structure Similarity Index Measure) between the original images (with acceptable quality) and their corresponding low resolution images (with unacceptable quality), since SSIM is partially related to image content. Table \ref{tab:data_iqa_summary} shows the mean and standard deviation of SSIM values for the participants ratings are equal to 0.63 and 0.06, respectively. Therefore, we use an SSIM value around 0.63 as a criterion to determine how much noise we should apply for each noise model. The generated noisy images are considered as the images with unacceptable quality and are included in the training of the classifier.

% As we went through the augmentation methods in the previous sections, we will discuss how the noise level is determined in this section. From the previous Figure \ref{fig:figure_gaussian} and Figure \ref{fig:figure_salt}, we know that the amount of noise, added into the original image for generating the image with unacceptable quality, depends on the content of the image. Therefore, SSIM (Structure Similarity Index Measure) is used for analysis since SSIM is partially related to image content.

\begin{table}[t]
    \caption{SSIM Statistics of Unacceptable and Augmentation Data.}
    \centering
    \begin{tabular}{|c|c|c|} 
        \hline 
        & Mean& Standard Deviation\\ \hline
        Unacceptable (human)&0.63&0.06\\ \hline
        Gaussian Noise (model)&0.59&0.001\\ \hline
        S\&P Noise (model)&0.62&0.002\\ \hline
        GAN-based (model) & 0.738& 0.006\\ \hline
    \end{tabular}

    \label{tab:data_iqa_summary}
\end{table}

% The SSIM values, between the original images (with acceptable quality) and their corresponding low resolutions image (with unacceptable quality), are calculated.  A statistical analysis is done on these SSIM values and the result is as follows. 

% For the SSIM values between original images and noisy images, we also calculate the mean and standard deviation.
% It is known that SSIM can be adjusted to any value if the noise level is controlled properly. However, the standard deviation is harder (or even impossible) to control while fixing the SSIM mean value. From Table \ref{tab:data_iqa_summary}, we observe that the standard deviation for the additive noise model is far less than the one in the original dataset. This allows us to understand that additive noise models alone are not enough to simulate all images of unacceptable quality. Therefore, we also use GAN to generate images of unacceptable quality. The SSIM standard deviation values, between original images and GAN generated images, is closer to the real situation compared to those generated by the  additive noise models.

% \section{Dataset Augmentation}
% \label{sec:dataaug}
% \subfile{sections/07_Data_Augmentation}

\section{Experimental Results}
\label{sec:result}
In this section, we provide the details of the experiment's setup, the results, and the discussion. We describe how the features introduced in Section \ref{sec:method} are combined and selected to achieve the best performance, as well as the evaluation metrics used including confusion matrices, precision, recall, and accuracy.

% The simulation results are discussed in this section. The explanation and discussion are composed of 2 parts.  The first part is to explain how these extracted features should be combined to achieve the best performance. The second part is to discuss the performance of the following metrics: confusion matrices, precision, recall and F-score.
\subsection{Experiment Setup}
We use 5-fold cross validation to quantitatively evaluate the proposed methods. For each subset of the data, we select $1/5$ for testing, and the rest for training. We report the performance by calculating the mean and variance accuracy among 100 runs of the 5-fold cross validation. Note that the augmented data are only used in training.
%\textcolor{red}{You need to describe how the experiments are set up. How the datasets are splitted for training, testing (and validation if any), did you do cross-validation, etc.}

\begin{table}[t]
    \caption{Mean and Variance for 5-Fold Average Classification Accuracy}
    \centering
    \begin{tabular}{|c|c|c|c|} 
        \hline 
        & Mean Accuracy & Variance Accuracy  \\ \hline
        Original & 90.82\% & 0.0012\% \\ \hline
        Augmented & 91.46\% & 0.0004\% \\ \hline
    \end{tabular}
    \label{tab:Accuracy}
\end{table}

\begin{table}[t]
\centering
\small
\caption{Precision, Recall and F1-Scores}
\begin{tabular}{|c|c|c|c|c|}           \hline 
\multirow{1}{*}{Setting} &\multirow{1}{*}{Class Type}    & {Precision} & {Recall} & {F1-Score}               \\   \hline 
\multirow{2}{*}{Original} & Acceptable& 0.92 & 0.973 & 0.944   \\   \cline{2-5}  
 & Unacceptable & 0.82 & 0.593 & 0.688           \\   \hline 
\multirow{2}{*}{Augmented} & Acceptable& 0.982 & 0.907 & 0.943   \\   \cline{2-5}  
 & Unacceptable & 0.67 & 0.92 & 0.775          \\   \hline 
% \multicolumn{2}{|c|}{Average}   & 76.42 & 66.52 &69.8 \\ \hline 
 \end{tabular}
 \label{tab:PR}
 \end{table}
\subsection{Classification Results}
Table \ref{tab:Accuracy} shows the results of classifying raster images in the scanned document as acceptable or unacceptable with and without augmented data in training. We observe that including the augmented data in training slightly improves the overall accuracy. However, when we take a closer look at the two settings as shown in Table \ref{tab:PR}, we noticed that the recall for unacceptable class under the original data setting is much lower, \textit{i.e.}, 0.593, compared to \textit{i.e.}, 0.92 when augmented data is included in the training. This is important since our goal is to determine whether or not a low resolution setting can be used for scanning the document. If we wrongly predict an image with unacceptable resolution as acceptable, the image would be stored at a resolution that is too low to be usable. On the other hand, if we predict an image with a relatively high resolution as ``Unacceptable", we may select a higher resolution and store the image at an increased file size. However, the image is still usable. 
% with data augmentation, the prediction for the ``Unacceptable"  images is improved significantly for the true negative rate, which increased from 0.593 to 0.922 when using the augmented dataset for training. 
The trade-off is that the true positive rate (recall for acceptable) dropped from 0.973 to 0.907 when augmented data is used. This implies that the model trained with augmented data tends to predict more images as unacceptable, including those images that were previously predicted as acceptable without the augmented data.

% We use standard metrics such as \textit{precision}, \textit{recall}, and \textit{accuracy} to evaluate the performance of our proposed method to classify raster images in the scanned document as acceptable or unacceptable for a given resolution. 
% Note that all 9 features are used here for fair comparison. 
% % \subsubsection{Original Dataset vs. Augmented Dataset}
% % For the performance metrics under the original dataset, the SVM classifier is only trained on the original dataset. The average accuracy is reported in Table \ref{tab:Accuracy}.  
% When we take a closer 
% To better understand the effect of including augmented data in training, 

% From Table \ref{tab:PR}, is shows that the precision and recall are 0.92 and 0.97, respectively. However, the true negative rate is much lower, \textit{i.e.}, 0.593. The poor performance is largely due to the data imbalance as more images are rated as ``Acceptable" compared to ``Unacceptable", as discussed in Section \ref{sec:dataanaysis}. Since our goal is to determine whether or not a low resolution setting can be used for scanned documents, true negative rate is more important. If we wrongly predict an image with unacceptable resolution as acceptable, the image would be stored at a resolution that is too low to be usable. On the other hand, if we predict an image with a relatively high resolution as ``Unacceptable", we may select a higher resolution and store the image at an increased file size. However, the image is still usable.
\begin{table}[t]
    \caption{Normalized Confusion Matrix for the Original Dataset and with augmented data training (in parentheses).}
    \centering
    \begin{tabular}{@{}cc cc@{}}
        \multicolumn{1}{c}{} &\multicolumn{1}{c}{} &\multicolumn{2}{c}{Predicted} \\ 
        %\cmidrule(lr){3-4}
        \multicolumn{1}{c}{} & 
        \multicolumn{1}{c}{} & 
        \multicolumn{1}{c}{Acceptable} & 
        \multicolumn{1}{c}{Unacceptable} \\ 
        \cline{2-4}
        \multirow[c]{2}{*}{\rotatebox[origin=tr]{90}{Actual}}
        & Acceptable  & 0.973 (0.907) & 0.027 (0.093)  \\[1.5ex]
        & Unacceptable  & 0.407 (0.078)  & 0.593 (0.922) \\ 
        \cline{2-4}
    \end{tabular}
    \label{tab:Original_Conf_Matrix}
\end{table}

\subsection{Feature Combinations}
% In this section, we discuss how the features are combined to optimize classification accuracy.  
As described in Section \ref{sec:method}, we use the sequential floating forward selection (SFFS) method to determine the set of most relevant features. Two different experiments are performed to determine the best combination. We first search the optimal combination of the features based on the original dataset, we then apply the same method to the augmented dataset.  

\subsubsection{Original Dataset vs. Augmented Dataset}
Table \ref{tab:feature_importance} shows the rankings of importance for the features using the SFFS method \cite{SFFS} on the original and augmented datasets in decreasing order. Suppose we use Table \ref{tab:feature_importance} to illustrate the different feature combinations. If we can only select one feature, then the Differential Spatial Activity (DSA) would be the choice because it gives the best performance with just a single feature.  The remaining features should be selected based on the order of their importance.  

% If we can select four features as the feature vectors for the classifier, then a combination of DSA, standard deviation of power spectrum difference, mean of power spectrum difference and standard deviation of edge density should be selected.

When comparing the difference between original dataset feature rankings and that with augmented dataset in Table \ref{tab:feature_importance} , we observe that edge related features, \textit{i.e.}, Differential Spatial Activity, Mean and StdDev of Edge Density, ranked lower when the augmented dataset is used, while the MSE-related features become more important. Such difference is likely contributed by the added noisy images in the augmented dataset, which leads to increased importance of the residual error features, especially those related to MSE. Also, by adding noise to an image, the edges become less sharp, causing the decrease in rankings for edge-related features.  

% the noisy images in the augmented dataset added to balance the dataset, especially the additive noise degraded dataset, such as the Gaussian noise and salt\& pepper noise, which is directly adding noise to the images. This will emphasize the importance of the residual error features, especially the MSE-related features. Also, since adding noise to an image will make the sharp edges not obvious, this causes the decrease in rankings for edge-related features.  
\begin{table}[!t]
\small
    \caption{Ranking of the Features According to Their Importance For the Original Dataset in Decreasing Order.}
    \centering
    \begin{tabular}{|c|c|c|c|} 
        \hline 
        Feature Name& Category & Original & Aug \\ \hline
        Differential Spatial Activity& Edge &1 &2\\ \hline
        StdDev of PSD& Residual Error & 2 &5\\ \hline
        Mean of PSD& Residual Error &3&7\\ \hline
        StdDev of Edge Density& Edge &4&8\\ \hline
        Mean of Edge Density& Edge &5&9\\ \hline
        StdDev of Tile-SSIM& Image Structure & 6&6\\ \hline
        StdDev of MSE& Residual Error & 7&3\\ \hline
        Mean of Tile-SSIM& Image Structure & 8&4\\ \hline
        Mean of MSE& Residual Error & 9&1\\ \hline
    \end{tabular}
    \label{tab:feature_importance}
\end{table}

\subsubsection{Classification Performance for Selected Features}
\label{classificationsubsec}
\begin{figure}[h]
\centering
    \includegraphics[width=0.8\linewidth]{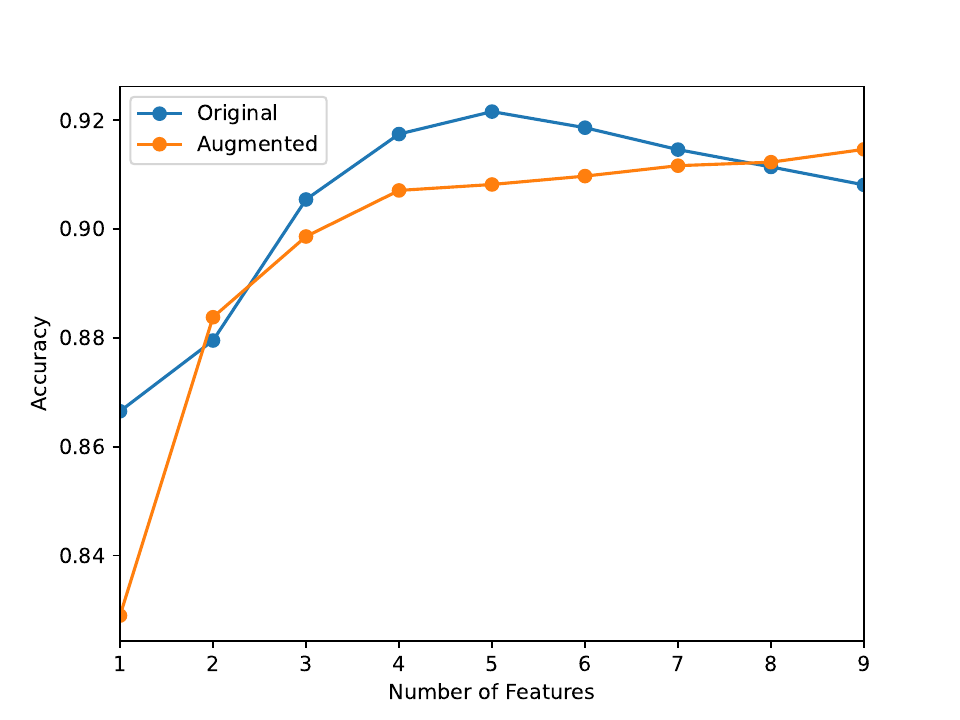}
   \caption{Classification Accuracy vs. Feature Selection.}
   \label{fig:fs_results}
\end{figure}

Figure \ref{fig:fs_results} shows the classification accuracy for different number of features used in the original dataset and the augmented dataset. 
% The average ranking for importance of features for each dataset can be found at Table \ref{tab:feature_importance} and Table \ref{tab:feature_importance_aug}.
We observe that a minimum of 3 features are needed to reach a classification accuracy above 90\% for the original dataset and 4 features for the augmented datasets. For the original dataset, the best performance is achieved when 5 features are selected, after which the performance drops slightly.  For the augmented dataset, the performance is quite stable when 4 or more features are selected. However, because the number of unacceptable images from the subjective ratings is very limited in the original dataset, it is necessary to include the additional noisy images in the augmented dataset to improve the performance of unacceptable images as discussed in Section \ref{sec:dataanaysis}. % and Section \ref{sec:result}.  

\vspace{-0.3cm}
\section{Conclusion}
\label{sec:conclusion}

In this work, we propose a low cost, efficient document image quality assessment method to estimate the visual quality of raster image content in scanned document pages under different resolution settings. We conduct a psychophysical experiment to collect human subject ratings of the visual quality of a variety of scanned documents. These ratings are used to train a machine learning model to determine whether the raster image in a scanned document page is visually acceptable or not at a given resolution. As the visual quality ratings are imbalanced for our dataset, we select three noise models to simulate the quality degradation during the scanning process and use them to augment the images in the original dataset. We leverage a feature selection strategy to determine the most relevant features to boost the performance of the classifier.  Our results show significant improvement by including augmented data in training for classification.

\printbibliography

@ARTICLE{Prima_Paper,
  author={Antonacopoulos, A. and Bridson, D. and Papadopoulos, C. and Pletschacher, S.},
  journal={Proceedings of the International Conference on Document Analysis and Recognition}, 
  title={A Realistic Dataset for Performance Evaluation of Document Layout Analysis}, 
  year={2009},
  month={July},
  pages={296-300},
  doi={10.1109/ICDAR.2009.271}}

@ARTICLE{litao_IQA,
  author={Hu, L. and Hu, Z. and Bauer, P. and Harris, T. J. and Allebach, J. P.},
  journal={Proceedings of the Electronic Imaging}, 
  title={Document Image Quality Assessment with Relaying Reference to Determine Minimum Readable Resolution for Compression}, 
  year={2020},
  month={January},
  pages={3231-3238},
  doi={10.2352/ISSN.2470-1173.2020.9.IQSP-323}}

@ARTICLE{litao_TextQA,
  author={Hu, L. and Hu, Z. and Bauer, P. and Harris, T. J and Allebach, J. P.},
  journal={Proceedings of the Electronic Imaging}, 
  title={Deep Learning Approaches to Determining Optimal Resolution for Scanned Text Documents}, 
  year={2021},
  month={January},
  pages={2431-2438},
  doi={10.2352/ISSN.2470-1173.2021.16.COLOR-243}}

@misc{wada2018labelme,
  author = {Kentaro Wada},
  title = {labelme: Image Polygonal Annotation with Python},
  year = {2018},
  publisher = {GitHub},
  journal = {GitHub repository},
  howpublished = {\url{https://github.com/wkentaro/labelme}},
  commit = {master},}

@article{canny1986computational,
  title={A computational approach to edge detection},
  author={Canny, J.},
  journal={IEEE Transactions on pattern analysis and machine intelligence},
  number={6},
  pages={679--698},
  year={1986},
  month={November},
  publisher={Ieee}
}

@article{DSA,
author = {Freitas, P. and Akamine, W. and Farias, M.},
year = {2018},
month = {May},
pages = {1-10},
title = {Using multiple spatio-temporal features to estimate video quality},
volume = {64},
journal = {Signal Processing: Image Communication},
doi = {10.1016/j.image.2018.02.010}
}

@ARTICLE{NIPS2014_5ca3e9b1,
 author = {Goodfellow, I. and Pouget-Abadie, J. and Mirza, M. and Xu, B. and Warde-Farley, D. and Ozair, S. and Courville, A. and Bengio, Y.},
 journal = {Advances in Neural Information Processing Systems},
 pages = {2672--2680},
 title = {Generative Adversarial Nets},
 volume = {27},
 year = {2014},
 month = {December}
}

@ARTICLE{tile_ssim,
  author={Hu, Z. and Hu, L. and Bauer, P. and Harris, T. J. and Allebach, J. P.},
  journal={Proceedings of the Electronic Imaging}, 
  title={Relation Between Image Quality and Scan Resolution: Part I}, 
  year={2020},
  month={January},
  pages={3221-3227},
  doi={10.2352/ISSN.2470-1173.2020.9.IQSP-322}}

@ARTICLE{DIQ_DL_Le,
  author={Kang, L. and Ye, P. and Li, Y. and Doermann, D.},
  journal={Proceedings of the International Conference on Image Processing}, 
  title={A deep learning approach to document image quality assessment}, 
  year={2014},
  month={October},
  volume={},
  number={},
  pages={2570-2574},
  doi={10.1109/ICIP.2014.7025520}}

@ARTICLE{DIQ_Kanungo,
  author={Kanungo, T. and Haralick, R.M. and Baird, H.S. and Stuezle, W. and Madigan, D.},
  journal={IEEE Transactions on Pattern Analysis and Machine Intelligence}, 
  title={A statistical, nonparametric methodology for document degradation model validation}, 
  year={2000},
  month={November},
  volume={22},
  number={11},
  pages={1209-1223},
  doi={10.1109/34.888707}}

@article{SVM,
  title={Support-vector networks},
  author={Cortes, C. and Vapnik, V.},
  journal={Machine learning},
  volume={20},
  number={3},
  pages={273--297},
  year={1995},
  publisher={Springer}
}

@article{SFFS,
title = {Floating search methods in feature selection},
journal = {Pattern Recognition Letters},
volume = {15},
number = {11},
pages = {1119-1125},
year = {1994},
month={},
issn = {0167-8655},
doi = {https://doi.org/10.1016/0167-8655(94)90127-9},
author = {P. Pudil and J. Novovičová and J. Kittler},
}

%%%%%%%%%%%%%%%%%%%%%%%%%%%%%%%%%%
% Biography
%%%%%%%%%%%%%%%%%%%%%%%%%%%%%%%%%%

% \begin{biography}
% Please submit a brief biographical sketch of no more than 75 words. 
% Include relevant professional and educational information as shown 
% in the example below.

% Jane Doe received her BS in physics from the University of Nevada (1977) 
% and her PhD in applied physics from Columbia University (1983). Since 
% then she has worked in the Research and Technology Division at Xerox 
% in Webster, NY. Her work has focused on the development of toner adhesion 
% and transport issues. She is on the Board of  IS\&T and a member of APS 
% and SPIE.
% \end{biography}

\end{document}